\documentstyle[twoside,fleqn,espcrc2]{article}

\newcommand{\be}{\begin{equation}}
\newcommand{\ee}{\end{equation}}
\newcommand{\bea}{\begin{eqnarray}}
\newcommand{\eea}{\end{eqnarray}}
\newcommand{\bc}{\begin{center}}
\newcommand{\ec}{\end{center}}

\def\ut{\tilde{U}}
\def\mt{\tilde{M}}

\title{\bf A new method for Monte Carlo simulation of theories
with Grassmann variables.}
\author{T. Bakeyev
\address{Joint Institute for Nuclear Research, 141980 Dubna, Russia}
       }

\begin{document}

\begin{abstract}

A new algorithm for simulation of theories with dynamical fermions is
presented.
The algorithm is based on obtaining the new configuration $\ut$
from the old one $U$ by solving the equation $ M(\ut) \eta= \omega M(U) \eta$,
where $M$ is fermionic operator,
$\eta$ is random Gaussian vector, and $\omega$ is random real number
close to unity.
This algorithm can be used for acceleration of current
simulations in theories with fermions.
\end{abstract}

\maketitle


From the earliest days of Monte Carlo simulations
fermionic fields have caused annoying difficulty,
which stems from their being anticommuting variables.
Although for the most actions in use one can eliminate
fermions by an analytic integration,
the resulting expressions involve determinants
of large matrices, making the numerical simulations expensive.

Over the years many tricks have been developed to solve this
problem. Hybrid Monte Carlo (HMC) \cite{HMC}
is now considered to be the standard simulation tool.
In recent years it
was challenged by the Multiboson (MB) method.
The most popular version of MB,
proposed by M.~L\"uscher \cite{Luscher:1994xx}, has been intensively studied by
many authors (see \cite{Lutest} and references therein).
The MB algorithm, proposed by
A.~Slavnov \cite{Slavnov}, is less known, and was tested in \cite{SLtest}.
Besides HMC and MB, let me mention
some interesting ideas, based on
the polymer \cite{polymer} and Jordan-Wigner occupation number \cite{OCCUP}
representations,
the direct evaluation of Grassmann integrals \cite{CR2},
the separation of low and high eigenmodes of Dirac
operator \cite{Infrared},
and
the direct simulation of loop expansion by means of using the stochastic
estimators \cite{NMC}.

Despite the variety of different approaches to the problem,
there is a common impression that all existing algorithms remain
inefficient \cite{Creutz}. This is clearly seen if we compare the
computational cost for the theories with dynamical fermions from the one side,
and purely bosonic theories from the other side.
Such a situation suggests that one should not stop the attempts to obtain
relatively cheap fermion algorithm.

In this contribution I present a new computational strategy for
treating dynamical fermions in Monte Carlo simulations.
It has the virtues of exactness, nonlocality and finite step-size of update
Unlike the other exact algorithms,
the CPU-time required for the update in this algorithm grows
only linearly with the volume $V$ of the system.

Suppose that we aim at sampling the partition function
of theory with two flavors of degenerate fermion fields:
\be {\cal Z}_{ferm} = \int Det \;\Bigl( M^\dagger [U] M[U] \Bigr)
\; dU \label{par_func}\ee
where $M$ is fermion operator, $U$ denotes the bosonic fields
coupled to fermions.
(For simplicity of formulae I do not include the purely
bosonic contribution.)
Then starting from old configuration $U$, one executes the following
instructions:

\begin{itemize}
\item generate random vector $\eta$ with Gaussian
distribution:
$ \;\;\; P_G[\eta] = Z^{-1}_G e^{-|\eta|^2} $
\item generate
$\;\omega \in \Bigl[1-\epsilon\; ;\; \frac{1}{1-\epsilon}\Bigr]\;$
with the probability:
${\cal P}[\omega,\eta]\propto min\; \bigl(1/\omega^2,
e^{\bigl(1 - \frac{1}{\omega^2}\bigr)|\eta|^2}\bigr)$
\item find the new configuration $\ut$ by solving the equation:
\be \mt \eta = \omega M\eta   \label{basic}\ee
\end{itemize}
where $0 \le \epsilon < 1$ is an algorithmic parameter.
Here and below I use the following short notations:
$\mt \equiv M[\ut] ;\; M \equiv M[U]\;$.

In Ref. \cite{Bakeyev:2001wb} it was proved that as far as the solution
of eq.(\ref{basic}) is found in a reversible way
\footnote{i.e. the probability to obtain
configuration $\ut$ starting from $U$ at any $\omega$ should be equal
to probability to obtain $U$ starting from $\ut$ at $1/\omega$},
the algorithm satisfies the detailed balance condition.

It is clear that choosing the procedure of solving
the eq.(\ref{basic}) is crucial in this approach. Firstly, because
obtaining the solution $\ut$ is the only
computationally expensive ingredient of the algorithm.
Secondly, because not every procedure of solving
the eq.(\ref{basic}) can provide ergodicity.
Finding a good procedure of solving the eq.(\ref{basic}) presents
a serious challenge for the future investigations. The
potential gain worth the efforts: e.g. finding
some analytic solutions can give us a very cheap fermionic algorithm,
comparable in cost with the algorithms for purely bosonic theories.

Making the first attempt to test the new algorithm, I solved
the eq.(\ref{basic}) numerically
by using the local iterative minimization of the quantity
\be R\equiv  |\; (\mt-\omega M)\eta \; |^2 \label{r_min}\ee
for fixed $U,\omega,\eta$. Starting from $\ut = U$,
the minimization proceeds until $R<\delta$ is reached, where $\delta$
determines the accuracy of solution.

The tests were done for SU(3) QCD at $8^4$ lattice for the
partition function (\ref{par_func}) with
\be M[U]=1 - k^2 D_{eo}D_{oe}\ee
where $\; D[U]$ is Wilson difference operator.
The even-odd preconditioning was used to reduce the computational cost.
The hopping parameter was $k=0.2$, which gave the plaquette
${\cal P} = 0.0089(1)$.
The minimization of functional (\ref{r_min}) was implemented by making
the random moves in each color direction for all links lexicographically.

I checked the existence of solutions of eq.(\ref{basic}) at
$\omega \in [0.9, 1.1]$ for typical equilibrated
configurations $U$. It was found that for considered
$\omega$ one always finds some solution.
On the average solving the eq.(\ref{basic}) at $\epsilon = 0.05$
with precision
$\delta = 5*10^{-7}$ required 42 minimization iterations
(it was checked, that improving the precision did not affect the results).
Since the cost of one iteration is roughly equal to $2*N_{generator}=16$
multiplications
by matrix $M$, the total cost of finding the solution
was approximately 670 matrix multiplication.
This has to be compared with the cost for generating one trajectory with HMC.
At step-size $\tau = 0.033$ and trajectory
length equal to $1$ (this ensures 70\% acceptance),
one trajectory cost approximately
4420 matrix multiplications, i.e. $\approx 6.6$ times more expensive than
solving eq.(\ref{basic}).

My conclusion was that the usage of minimization
of functional (\ref{r_min}) for solving the eq.(\ref{basic})
is not favorable, because the algorithm became nonergodic.
The problem is that probability
${\cal P}[\omega,\eta]$ to accept $\omega<1$ is
strongly suppressed by large factor in the exponent, if $\omega$
is far from 1.
Indeed, the squared norm of Gaussian vector can be estimated as
$\; |\eta|^2 \sim I_{dof}$, where $ I_{dof} $ is the dimensionality
of vector space on which $\eta$ is defined.
For even-odd preconditioned SU(3) QCD at $8^4$ lattice one has
$I_{dof} = 24576$!
Therefore, using the new algorithm alone,
one is restricted to choose between two unfavorable possibilities:
1) Using $\epsilon \sim 1/I_{dof} $.
It was observed that at such small values of $\epsilon$
the evolution of $U$ fields becomes
very slow, because the new configuration $\ut$ always lies too close to
the old one;
2) Using $\epsilon \gg 1/I_{dof}$. Then $\omega < 1$ is almost never
accepted, and the configurations are becoming smother.

In the second case the trouble appears due to the nonergodicity of algorithm
at large $\epsilon$. Nevertheless, even at large $\epsilon$
our algorithm can be used for acceleration of other fermionic
algorithms like HMC or MB, since the ergodicity is provided
by them.

I tested the combination of the new algorithm and HMC. After each
HMC trajectory the global move in configuration space was
implemented
by using our algorithm at $\epsilon = 0.05$.
The efficiency of this algorithmic
mixture was estimated by measuring
the autocorrelation times for the plaquette. The results were:
1) Pure HMC, 1000 trajectories: $\tau_{int}=1.4(2)$;
2) HMC+new alg., 1000 (traj + $\omega$-update): $\tau_{int}=0.7(1)$.
Note, that autocorellation time was reduced almost at no cost,
because the global update of the new algorithm is much cheaper, than HMC
trajectory.
Of course, these tests are not very illustrative, since the
autocorrelation times for theory with pure fermionic term at this kappa
value are rather small.
The further tests in models of practical relevance are needed.

Let me emphasize, that nonergodicity of the new algorithm
was produced by the particular way of solving the eq.(\ref{basic}),
which forces us to use large $\epsilon$.
One should think of finding some other procedure,
which allows to use $\epsilon = 0$ (or $\epsilon \approx 0$).
Unfortunately, in this case the search for the solution of eq.(\ref{basic})
by minimizing the functional (\ref{r_min}) does not work,
because one immediately gets the trivial solution $\ut = U$.
However, in typical case (e.g. SU(3) QCD), the solutions of eq.(\ref{basic})
are highly degenerate
\footnote{
In SU(3) QCD the expression (\ref{basic}) for even-odd preconditioned
matrix provides $12V$ real equations for $32V$ variables.},
and the subspace of nontrivial solutions is not empty.
Finding the reversible procedure
of nontrivial solving the equation $\mt\eta = M\eta$,
which samples the configuration space fast enough,
is the perspective subject for future investigations.

I conclude that the efficiency of the new fermion algorithm 
drastically 
depends on the clever choice of procedure of solving the 
eq.(\ref{basic}). Finding the new efficient way of solving this 
equation can provide a very cheap simulation method.

{\bf Acknowledgments:}
I am very grateful to the Organizers of LATTICE2001
for generous help and support.
I thank Ph. de Forcrand for helpful comments.
This work was supported by grant for young scientists
\# 01-01-06089, Russian Basic Research Fund
\# 99-01-00190 and president grant for support of
scientific schools \# 0015-96046.

\end{document}